\documentclass[11pt]{article}
\pdfoutput=1
\usepackage{epsf}
\usepackage{cite}
\usepackage{amsmath,amssymb,dsfont}
\usepackage[dvips]{graphicx}
\usepackage{comment}
\usepackage{color}
\usepackage{hyperref}

\newcommand{\lp}{\left(}
\newcommand{\rp}{\right)}
\newcommand{\ls}{\left[}
\newcommand{\rs}{\right]}

\begin{document}

\newcommand{\lsim}{\stackrel{<}{_\sim}}
\newcommand{\gsim}{\stackrel{>}{_\sim}}
\newcommand{\rem}[1]{{ {\color{red} [[$\spadesuit$ \bf #1 $\spadesuit$]]} }}
\numberwithin{equation}{section}


\def\thefootnote{\fnsymbol{footnote}}
\def\a{\alpha}
\def\b{\beta}
\def\c{\varepsilon}
\def\d{\delta}
\def\e{\epsilon}
\def\f{\phi}
\def\g{\gamma}
\def\h{\theta}
\def\k{\kappa}
\def\l{\lambda}
\def\m{\mu}
\def\n{\nu}
\def\p{\psi}
\def\q{\partial}
\def\r{\rho}
\def\s{\sigma}
\def\t{\tau}
\def\u{\upsilon}
\def\v{\varphi}
\def\w{\omega}
\def\x{\xi}
\def\y{\eta}
\def\z{\zeta}
\def\D{\Delta}
\def\G{\Gamma}
\def\H{\Theta}
\def\L{\Lambda}
\def\F{\Phi}
\def\P{\Psi}
\def\S{\Sigma}
\def\me{\mathrm e}

\def\o{\over}
\def\beq{\begin{eqnarray}}
\def\eeq{\end{eqnarray}}
\newcommand{\vev}[1]{ \left\langle {#1} \right\rangle }
\newcommand{\bra}[1]{ \langle {#1} | }
\newcommand{\ket}[1]{ | {#1} \rangle }
\newcommand{\bs}[1]{ {\boldsymbol {#1}} }
\newcommand{\mc}[1]{ {\mathcal {#1}} }
\newcommand{\mb}[1]{ {\mathbb {#1}} }
\newcommand{\EV}{ {\rm eV} }
\newcommand{\KEV}{ {\rm keV} }
\newcommand{\MEV}{ {\rm MeV} }
\newcommand{\GEV}{ {\rm GeV} }
\newcommand{\TEV}{ {\rm TeV} }
\def\diag{\mathop{\rm diag}\nolimits}
\def\Spin{\mathop{\rm Spin}}
\def\SO{\mathop{\rm SO}}
\def\O{\mathop{\rm O}}
\def\SU{\mathop{\rm SU}}
\def\U{\mathop{\rm U}}
\def\Sp{\mathop{\rm Sp}}
\def\SL{\mathop{\rm SL}}
\def\tr{\mathop{\rm tr}}
\def\sp{\;\;}

\def\IJMP{Int.~J.~Mod.~Phys. }
\def\MPL{Mod.~Phys.~Lett. }
\def\NP{Nucl.~Phys. }
\def\PL{Phys.~Lett. }
\def\PR{Phys.~Rev. }
\def\PRL{Phys.~Rev.~Lett. }
\def\PTP{Prog.~Theor.~Phys. }
\def\ZP{Z.~Phys. }

\def\K{K{\"a}hler}
\newcommand{\rf}[1]{(\ref{#1})}
\newcommand{\vp}{\varphi}

\def\draftnote#1{{\color{red} #1}}
\def\bldraft#1{{\color{blue} #1}}

\def\be{\begin{equation}}
\def\ee{\end{equation}}
\def\ba{\begin{array}}
\def\ea{\end{array}}

\newcommand{\mk}{\mathcal{K}}
\newcommand{\ml}{\mathcal{L}}
\newcommand{\mr}{\mathcal{R}}
\newcommand{\vt}{\vartheta}

\parskip 7pt


\begin{titlepage}

\begin{center}

\hfill CERN-TH-2016-156 \\
\hfill TUW-16-12\\

\vskip .75in

{\LARGE \bf 
 Sneutrino Inflation with $\alpha$-attractors
}

\vskip .75in

{\large Renata Kallosh$^{a}$, Andrei Linde$^{a}$, Diederik Roest$^{b,c}$ and Timm Wrase$^d$}

\vskip 0.25in

{\normalsize\sl\noindent
$^{a}$ {\em Stanford Institute of Theoretical Physics and Department of Physics, }\\
 {\em Stanford University, Stanford, CA, 94305 USA 
}\\[.3em]
$^{b}$ {\em Van Swinderen Institute for Particle Physics and Gravity,}\\
{\em University of Groningen, Nijenborgh 4, 9747 AG Groningen, The Netherlands}\\ [.3em]
$^c$ {\em Theoretical Physics Department, CERN, CH-1211 Geneva 23, Switzerland}\\[.3em]
$^d$ {\em Institute for Theoretical Physics, TU Wien, A-1040 Vienna, Austria}}

\end{center}
\vskip .5in

\begin{abstract}
Sneutrino inflation employs the fermionic partners of the inflaton and stabilizer field as right-handed neutrinos to realize the seesaw mechanism for light neutrino masses. We show that one can improve the latest version of this scenario and its consistency with the Planck data by embedding it in the theory of cosmological $\alpha$-attractors. 
\end{abstract}
\end{titlepage}


 \renewcommand{\thepage}{\arabic{page}}
 \setcounter{page}{1}
 \renewcommand{\thefootnote}{\#\arabic{footnote}}
 \setcounter{footnote}{0}


\section{Introduction}
\setcounter{equation}{0}

Inflation in the early universe is well described by the general class of $\alpha$-attractor models. These can be constructed in supergravity as F-term models \cite{Kallosh:2013yoa} or as D-term models \cite{Ferrara:2013rsa}, where the first $\alpha$-dependent generalization of the Starobinsky  potential was introduced. A robust feature of these models is the spectral index $n_s\approx 1-{2\over N}$, which is in perfect agreement with the observed value of $n_s$ for the number of $e$-foldings $50 \lesssim N \lesssim 60$, as one can see  in Fig. 12 in \cite{Ade:2015lrj}. 

The level of gravity waves in $\alpha$-attractors is given by  $r\approx  {12\alpha / N^2}$, so that $r\approx 4 \alpha \cdot 10^{-3}$ at $N\approx  55$. It is directly related to the curvature of the moduli space ${\cal R}_K= -{2/ (3 \alpha)}$ which in ${\cal N}=1$ supergravity  is a free parameter  \cite{Ferrara:2013rsa}. From the current Planck-BICEP-Keck constraint  $r< 0.07$ \cite{Ade:2015lrj,Array:2015xqh}  it follows that  $\alpha \lesssim 17$. Future B-mode experiments will constrain this parameter further. More precise constraints on $\alpha$  depend on the details of the models and on the mechanism of reheating, see e.g.   \cite{Ellis:2015pla,Kallosh:2016gqp}.

Inflation in supergravity is often described using two chiral superfields, the inflaton $T$ and the stabilizer $S$, as first suggested in \cite{Kawasaki:2000yn} for quadratic chaotic inflation, and developed for a general class of models in \cite{Kallosh:2010ug}. There are 4  real scalar fields in these models, one of which is the relatively light inflaton and 3 others vanish during inflation. There are situations where some of these 3 fields may acquire tachyonic masses, which may lead to instability. Fortunately, there are well established methods either to control these instabilities by adding stabilizing terms to the \K\ potential \cite{Kallosh:2010ug}, or to make these 3 extra fields disappear in the context of the theory of nilpotent orthogonal superfields   \cite{Ferrara:2015tyn,Carrasco:2015iij,Dall'Agata:2015lek}. 

In realistic theories of elementary particles one may encounter more than two chiral superfields, which again requires an investigation of stability during and after inflation. There are some general conditions which allow to avoid tachyonic instability for matter fields \cite{Kallosh:2016ndd}. These conditions are easy to satisfy if the fields $S$ and $T$ belong to a hidden sector and do not have direct interactions with matter fields in the \K\ potential and in the superpotential  \cite{Kallosh:2016ndd}. However, some popular inflationary models do not belong to this class. For example, Higgs inflation in supergravity is supposed to be driven by the Higgs field, which does not belong to the hidden sector. This may result in  instabilities during and after inflation. However, all of these instabilities can be successfully controlled \cite{Ferrara:2010yw}.

In this paper we will study sneutrino inflation \cite{Murayama:1992ua,Nakayama:2016gvg}, which is yet another scenario where the stability conditions found in \cite{Kallosh:2016ndd} may be violated \cite{Nakayama:2016gvg}, which typically results in the post-inflationary tachyonic instability.  
We should note from the very beginning that it is not the goal of our paper to formulate a fully consistent cosmological model of sneutrino inflation complemented by the theory of reheating, leptogenesis, SUSY breaking and many other aspects of this scenario closely related to particle phenomenology. These aspects have been studied in multiple papers on sneutrino inflation, see e.g. a discussion of reheating in \cite{Nakayama:2016gvg}, leptogenesis in \cite{Bjorkeroth:2016qsk}, and SUSY breaking in \cite{Linde:2016bcz}. Here we will limit ourselves to the investigation of inflationary aspects of this scenario as compared to the latest Planck data  \cite{Ade:2015lrj,Array:2015xqh}. In particular, we will show that one can improve consistency of sneutrino inflation with the latest Planck data by embedding it into the theory of $\alpha$-attractors. We will also develop a new approach to matter field stabilization in this scenario, which renders the post-inflationary tachyonic instability in this scenario harmless.

The latest version of the sneutrino inflation  \cite{Nakayama:2016gvg} refers to an extension of the model \cite{Kallosh:2010ug} with the neutrino sector, such that the fermions in the inflaton and stabilizer superfields provide the right-handed partners to the Standard Model neutrinos.  Then one generates light neutrino masses via the seesaw mechanism \cite{seesaw}. An interesting observation of \cite{Murayama:1992ua,Nakayama:2016gvg} is that in large field models of inflation the inflaton mass is of order $M\sim 10^{13}$ GeV, which is close to the right-handed neutrino mass scale suggested by the seesaw mechanism and the neutrino oscillation experiments. Explicitly, the left handed neutrino masses come from 
 integrating out  the heavy fermions with mass $M$ from the  inflationary sector,  so that
 \be
m_{\rm light}^\nu \sim  {v^2\over M} \sim {10^4\over  10^{13} }{\rm \ GeV}   \sim {\rm \  eV}\,,
 \ee 
where $v$ is  the Higgs vev.  The left-handed neutrinos are many orders of magnitude lighter than all the other particles in the Standard Model and the seesaw mechanism is the only known way to explain this, but it does not give a reason for the existence of the heavy fermions  of mass $\sim 10^{13}$ GeV. Inflation seems to naturally supply these heavy fields at the right scale. In particular, the general supergravity constructions  \cite{Kawasaki:2000yn,Kallosh:2010ug} necessarily involve two superfields (the inflaton and the stabilizer fields) and thus provide two right-handed neutrinos, allowing for two massive and one massless neutrino.   

The  model of sneutrino inflation proposed in \cite{Nakayama:2016gvg} is based on the supergravity implementation of the generalized natural inflation scenario   proposed in \cite{Kallosh:2014vja} with the superpotential \footnote{The authors of  \cite{Nakayama:2016gvg} call it multi-natural inflation \cite{Czerny:2014wza}, but the supergravity versions of the multi-natural inflation models proposed in \cite{Czerny:2014wza} are quite different; they involve axion superfields without the stabilizer superfield $S$.} 
 \be
W= M S  \sum_n {g_n \over n} f \sin {\lp n \varphi\over f\rp} \,.
\ee
 The simple natural inflationary model  with a single frequency  is disfavored by the recent Planck data  \cite{Ade:2015lrj,Array:2015xqh}. Moreover, when adding a second frequency term in the superpotential to fit the data, it was observed in \cite{Nakayama:2016gvg} that  its coefficient must  have a complex coefficient $g_2= Ce^{i\theta}$. Models with complex parameters in the superpotential are somewhat complicated, as shown in \cite{Kallosh:2014xwa}. To fit the future data on $n_s$ and especially $r$ will require a fine tuned choice of $f, C, \theta$ parameters. This is certainly possible but will become more difficult when the bound on $r$ will decrease further, as one can see from Fig. 2 in  \cite{Nakayama:2016gvg} where various numerical examples are plotted.

The important feature of the multi-frequency natural inflation models which allows them to be used for sneutrino inflation is the unbroken discrete shift symmetry   $\vp \rightarrow \vp + 2\pi f$. This symmetry allows one to keep the inflationary predictions intact when 4 additional superfields, 3 leptons $L^a$, $a=1,2,3$ and the Higgs $H^u$ are added to the model via the following superpotential \cite{Nakayama:2016gvg}
\be
W=  p_a  L^a H^u \sum_n {g'_n \over n} f \sin {\lp n \varphi\over f\rp}\,.
\ee
This is invariant under $\vp \rightarrow \vp + 2\pi f$, thus even at large $\vp$ the masses of the  lepton and Higgs supermultiplets are not blowing up since $\sin {\lp n \varphi /  f\rp}$ is restricted  for any value of $\vp$. 
 
However, the $\alpha$-attractor models do not have an unbroken discrete shift symmetry; they have instead a symmetry associated with the hyperbolic geometry. Will these models provide a  framework for sneutrino inflation models? The answer to this question is positive, as we will argue below.

In hyperbolic $\alpha$-attractor models,  the leptons and the Higgs field interact with the disk variable $Z=\tanh  {\lp \Phi/ \sqrt 6 \alpha\rp} $. In the simplest case we have
\be
W =  p_a  \,  L^a \, H^u \, Z =  p_a \, L^a \, H^u \,  \tanh {\lp \Phi \over  \sqrt {6 \alpha}\rp} \,,
\ee
analogous to a single Fourier term in  natural inflation. Inflation occurs when $Z$ is very close to the boundary of the moduli space at $Z \bar Z \approx 1$, which corresponds to $\vp \gg   \sqrt 6 \alpha$ in terms of the canonically normalized inflaton field $\vp \equiv$Re$(\Phi)$. For a \K\ potential that describes hyperbolic geometry supplemented with a generic superpotential that is non-singular at the boundary of the moduli space, the inflaton potential $V(\vp)$ becomes flat. In this limit, the masses of all particles at large values of the inflaton field $\vp$ approach exponentially fast their constant values corresponding to the limit $Z\to 1$ or $\vp \to \infty$, and all coupling constants of the field $\vp$ to the matter fields become exponentially small in the limit $\vp \gg  \sqrt 6 \alpha$
\cite{Kallosh:2016ndd,Kallosh:2016gqp}.

Thus the hyperbolic geometry of $\alpha$-attractors protects the masses of leptons and Higgs fields during inflation, and suggests inflationary models which are more flexible with regard to future data and have an elegant geometric interpretation. Instead of periodic functions of the inflaton $\sin {\lp n \varphi / f\rp}$ interacting with matter multiplets, we have a natural geometric variable of the moduli space with a boundary, the coordinate of the Poincar\'e disk $Z= \tanh {\lp \Phi / \sqrt{6 \alpha}\rp}$ where $Z\bar Z<1$. An interpretation of this moduli space in terms of Escher's picture Circle Limit IV was proposed in \cite{Kallosh:2015zsa}, where it was shown that the radius square of this disk is given by $3\alpha$.
Thus the trigonometric restriction on couplings to matter is replaced by the hyperbolic one:
\be
\sin  {\lp n \varphi\over f\rp} \leq 1 \,  \qquad   \text{is replaced by} \qquad Z\bar Z  < 1\,,
\ee
while the role of the axion decay constant $f$ is played by the curvature parameter $\alpha$. This suggests that one may consistently implement sneutrino inflation in the context of the theory of $\alpha$ attractors. 

An advantage of implementing inflation in the theory of $\alpha$-attractors becomes obvious when one recalls that the consistency of the model of Ref. \cite{Nakayama:2016gvg} with the Planck data requires the fine-tuning of 2 parameters, $f$, and $\theta$, see Fig. 2 in  \cite{Nakayama:2016gvg}. Meanwhile  the predictions of $\alpha$-attractors are compatible with observational data for all $\alpha \lesssim O(10)$, without any fine-tuning. 

To verify the consistency of the sneutrino inflation scenario  one should check the stability of the system with respect to the generation of large vevs of the many scalar fields involved. Indeed, in this  scenario the interactions of the inflaton field with matter  do not satisfy the stability conditions formulated in \cite{Kallosh:2016ndd}. It was argued in \cite{Nakayama:2016gvg} that the stability of the inflationary regime in this scenario can be achieved by adding stabilizing terms to the \K\ potential. We confirm this conclusion, but find that unless the coefficients in front of these terms are incredibly large, the tachyonic instability does appear after the end of inflation. Adding the stabilizing terms to the \K\ potential does not help here; moreover, once the instability starts, these terms make the development of the instability unpredictable.  Fortunately, we found a simpler way to achieve matter field stabilization during inflation, without introducing higher order terms in the \K\ potential. Even with this novel stabilization mechanism, the tachyonic instability does appear after inflation, but we found that in our scenario this instability is transient and harmless. It is just a part of a post-inflationary tachyonic preheating \cite{Felder:2000hj}, similar to the waterfall regime at the end of the hybrid inflation scenario \cite{Linde:1991km}. 

Our results may have implications for  general models of multi-field inflation, going beyond sneutrino inflation. In multi-field models, the evolution of scalar fields may go along a very complicated path in a random potential, which may render long and stable large-field inflationary trajectories unlikely, see e.g. \cite{Aazami:2005jf,Marsh:2013qca,Freivogel:2016kxc}. The situation improves if one of these fields belongs to the class of $\alpha$ attractors, which requires it to have a singular kinetic term at some point in the moduli space \cite{Kallosh:2013daa}. Indeed, it is easier to avoid instabilities in a vicinity of a single point. Then, upon transition to canonical variables, the potential acquires an infinitely long  stabilized flat inflaton direction. At large values of the inflaton field along this direction, it essentially decouples from all matter fields, making them irrelevant for the development of inflation \cite{Kallosh:2016gqp}. When inflation along this direction ends, all other fields may fall down in a non-inflationary manner without affecting the main inflationary predictions and creating the problems described in \cite{Aazami:2005jf,Marsh:2013qca,Freivogel:2016kxc}. 

The main remaining problem is to implement these ideas in supergravity and  make sure that these fields do not fall to a collapsing vacuum state with a negative vacuum energy. This problem does not appear, if the theory has a stable supersymmetric Minkowski vacuum in which it settles at the end of inflation. We will show that this is indeed the case in the sneutrino inflation model that we analyze. Of course, this is not the end of the story, since our vacuum is not supersymmetric, but this is not a real problem either, because one can  break supersymmetry and uplift the vacuum in a way that does not affect the inflationary evolution, see e.g. \cite{Linde:2016bcz} and references therein. This suggests that the theory of cosmological attractors in combination with vacuum stabilization may significantly simplify the construction of realistic multifield inflationary models in supergravity.

The outline of this paper is as follows. In section 2 we will provide a review of $\alpha$-attractor models. These will be coupled to the neutrino sector in section 3. The stability during and after inflation and the role of tachyonic directions is discussed in section 4. We close with a discussion and present our conclusions in section 5.

\section{Inflationary models based on hyperbolic geometry}

The key aspect of $\alpha$-attractors is their hyperbolic geometry for the moduli space spanned by the scalar fields. This hyperbolic geometry is often parametrized as the Poincar\'{e} half-plane or disk. We will focus on the latter formulation in this paper. In the disk  variables our models are\footnote{In half-plane variables the same models become
\be
K= -{3 \alpha \over 2}  \log \left[ {(T + \bar T )^2 \over  4 T \bar T}   \right] + S\bar S\,  ,  \qquad W= G(T)+ S F(T)\ ,\nonumber
\ee
related via the Cayley transformations $T= {1+Z\over 1-Z}$, $ Z= {T-1\over T+1}$.}
\be
K= -{3\alpha\over 2}    \log \left[{(1- Z\bar Z)^2\over (1-Z^2) (1-\overline Z^2)}  \right] +S\bar S\,,  \qquad W= A(Z)  + S B(Z)\,   \,,
\label{KdiskNew}\ee
with the superpotential containing two arbitrary but real holomorphic functions.\footnote{A real holomorphic function $f(\Phi)$ is a holomorphic function such that $\bar f(\bar\Phi) = f(\Phi)$.} 
The K\"{a}hler metric for these models reads
\be
K_{Z\bar Z} = \frac{3\alpha}{(1-Z\bar Z)^{2}} \,.
\ee
During inflation we have $S=\bar S=0$ and $Z=\bar Z$ so that the inflaton is ${\rm Re} (Z)$, and  
\be
\qquad K=0\, ,  \qquad K_Z=0\,.
\ee
For our purpose of providing a pair of heavy fermions from inflationary multiplets, we will not use constrained superfields, but just assume that we can add some stabilizing features \cite{Kallosh:2010ug} to our \K\, potential. For example, we can add terms like $(S\bar S)^2$ and $ S\bar S  (Z-\bar Z)^2$ leading to sectional and bisectional curvatures which stabilize the 3 scalars other than Re$(Z)$, so that only the inflaton is relatively light. The potential during inflation is given by
\be
V= {1\over 3\alpha} A(Z)'^2 (1-Z^2)^2 + B(Z)^2 -3 A(Z)^2\,,
\ee
evaluated along the real part of $Z$.

An alternative parametrization of the hyperbolic geometry is in terms of the Killing coordinates
\be
\Phi= \vp +i \vt\, , \qquad 
Z= \tanh {\lp \Phi\over \sqrt {6 \alpha}\rp}\, , \qquad 
T= e^{\sqrt{2\over 3\alpha} \Phi} \,.
\ee
This change of variables leads to
 \be
 K= -3\alpha \log \ls\cosh {\lp \Phi-\bar \Phi \over \sqrt{6\alpha}\rp} \rs + S \bar{S} \,,
\ee
and the superpotential becomes 
\be
 W= A \lp\tanh  {\lp\Phi \over \sqrt {6\alpha}\rp}  \rp  + S\, B \lp\tanh {\lp\Phi \over \sqrt {6\alpha}\rp}\rp\equiv g(\Phi)+ S f(\Phi)\,,
\ee
where $f$ and $g$ are arbitrary real holomorphic functions.
The metric is conformally flat for all parametrizations 
\be
ds^2=  3\alpha {dZ d\bar Z\over (1-Z\bar Z)^2}= 3\alpha {dT d\bar T\over (T+\bar T)^2}= {\partial \Phi \partial \bar \Phi\over 2 \cosh^2\Big ((\Phi - \bar \Phi) / \sqrt{6\alpha}\Big )}\,  .
\ee
The special feature of the Killing coordinates is that the metric is independent of the inflation (which is the real component in all cases); indeed, along $\Phi=\bar \Phi = \vp$, the metric for the inflaton  is canonical. Finally, the corresponding scalar potential takes the particularly simple form
\be
V_{\rm total}=  2 g'(\vp)^2 - 3  g(\vp)^2 +  f(\vp)^2 \,.
\ee
For all choices of moduli space coordinates above, both the  \K\, potential and its first derivative vanish during inflation, making the \K\, covariant derivatives identical to simple derivatives $D_\Phi W = \partial _\Phi W=g'(\Phi)$,  $D_S W= \partial_S W= f(\Phi)$.  In the case of Killing coordinates, the additional simplification is that the inflaton's inverse kinetic term is simple,  $K^{\Phi\bar \Phi}=2$.  
By a proper choice  of the model, one can achieve a desirable value for the cosmological constant and the SUSY breaking scale. This can be done in a most natural way using the theory of constrained superfields, see e.g. \cite{Ferrara:2015tyn,Dall'Agata:2015lek,Carrasco:2015iij}.

The simplest example of a model based on hyperbolic geometry, analogous to the first supergravity model for quadratic inflation with a flat \K\, geometry  \cite{Kawasaki:2000yn}, is given by
\be
W=  M S Z \qquad \Rightarrow \qquad 
V=   M^{2}Z^{2} =   M^2 \tanh^2{\lp \vp\over \sqrt {6\alpha}\rp} \ .
\label{KdiskNew2}\ee
During inflation, which takes place close to the boundary $Z=1$, the potential is very close to the plateau $V = M^2$ and the Hubble constant is given by $H^{2} = M^2/3$. A proper COBE-Planck normalization of the scalar perturbations is achieved by the choice \cite{Kallosh:2015lwa}
\be
M \sim 10^{{-5}}\, \sqrt\alpha \ .
\ee
Instead, near the minimum of the inflaton potential at $Z = 0$ (i.e for $\vp \ll \sqrt{6\alpha}$), the scalar potential reduces to $M^{2}\vp^{2}/(6 \alpha)$.  The mass of the inflaton field is then given by $M/\sqrt{3\alpha}$.

An investigation of this model during inflation in the slow-roll approximation yields the following field values at the moment when the universe was $N$ e-foldings away from the end of inflation: 
\be
\vp_{N} = \sqrt{3\alpha\over 2} \log{\lp8N\over 3 \alpha\rp} \,, \quad Z_{N} =  \tanh\ls{1\over 2}\log{\lp 8N\over 3\alpha\rp}\rs \ .
\ee
Note that $Z_{N}$  approaches $1$, the boundary of the disk, in the limit $N \to \infty$.  For future reference we will give here some numerical results for the simplest case $\alpha = 1$: 
 \begin{align}
  N = \{ 50, 10, 5, 2, 1 \} \quad \Rightarrow \quad \begin{cases} \quad Z_N = \{ 0.985, 0.928, 0.860, 0.684, 0.455 \} \,, \\ \quad
 \vp_N = \{ 5.99, 4.02, 3.17, 2.05, 1.20 \} \,. \end{cases} 
 \end{align}
We will return to these numerical values later, since the effective masses of all other fields during and after inflation will be directly related to the field $Z$.

It should be noted that the field values quoted at small $N$ are not very precise because of the failure of the slow-roll approximation at the end of inflation and also because there are several slightly inequivalent definitions of the ``end of inflation''. One popular definition is that it is the moment when one of the slow roll parameters becomes equal to $1$, yet another one is that the end of inflation is the moment when the universe no longer accelerates. Using this last definition and solving the equations without using the slow-roll approximation gives the value of the field $Z$ at the end of inflation $Z_{0} \sim 0.33$. In addition, all field values quoted here are model-dependent: in contrast to the inflationary predictions around $N=50$, where the specific choice of the functions in the superpotential only affects subleading effects in $1/N$, the behaviour for small values of $N$ can differ significantly between models.

\section{Adding the neutrino sector}

The next step is to add to the model \rf{KdiskNew} the {\K} and superpotential for the MSSM fields. The RH neutrinos are the fermions in the $Z$ and $ S$ multiplets.  In addition to the inflationary part, the superpotential has some terms where the leptons and Higgs multiplets interact with the inflationary sector,
\beq
K&=& -{3\over 2}   \alpha \log \left[{(1- Z\bar Z)^2\over (1-Z^2) (1-\overline Z^2)}  \right] +S\bar S + \delta_{a\bar a} L^a \bar L^{\bar a} + H^u \bar H^u\,,\cr
W&=& A(Z)+ S B(Z) + (p_a \, Z + q_a \, S )  L^a H^u \, .
\label{KhalfNew}
\eeq
There are 6 complex parameters $p_a$ and $q_a$, $a=1,2,3$ which serve to build the light neutrino mass matrix upon integrating out the RH neutrinos. 

Recently the generic coupling of models of inflation to a matter sector was investigated in \cite{Kallosh:2016ndd}. The criteria spelled out there, that in particular lead to the absence of tachyons in the matter sector during and after inflation, are not satisfied for the generic model in eqn. \eqref{KhalfNew}. This means that generically one expects that some of the matter fields become tachyonic at a certain stage of the inflationary evolution, which is indeed the case as we will see below. 

Since we cannot rely on generic criteria that ensure the stability of the matter fields, we have to do an explicit analysis of the full model given in eqn. \eqref{KhalfNew}. In order to do so let us note some important points about models of the form given in eqn. \eqref{KhalfNew}, which follow from explicit calculations as carried out for example in \cite{Kallosh:2016ndd}: There is a critical point for the matter fields at $L^a=H^u=0$ and at this critical point the matter sector does not affect the inflationary sector at all since mixed derivatives of the potential involving $Z$ or $S$ and one of the matter fields $L^a$ or $H^u$ vanish. We thus can focus on the mass matrix for the matter fields at this critical point during and after inflation. 

In the Standard Model, $L^a$ and $H^u$ are electro-weak $SU(2)$ doublets, while $L^a H^u$ is the invariant inner product. However, for the above {\K} and superpotential one finds that the $16\times16$ matter sector mass matrix at $L^a=H^u=0$, which will be satisfied during inflation, factorizes into two identical $8 \times 8$ blocks that involve either only first or only second components of both doublets. So there is no mixing between both components and they have the same $8\times8$ mass matrix. This $8\times 8$ mass matrix can be simply obtained by treating $L^a$ and $H^u$ effectively as $SU(2)$ singlets, which is what we are going to do in this section. However, it is important to keep in mind that they are doublets and that the full $16 \times 16$ mass matrix only factorizes into two decoupled and identical $8\times 8$  blocks if $L^a=H^u=0$.

In order to calculate this mass matrix for the four complex fields $L^a$ and $H^u$ it is useful to first take advantage of the symmetries of our model. The {\K} potential part $\delta_{a\bar a} L^a \bar L^{\bar a}$ is invariant under $SU(3)$ transformations acting on the $a$ index. This allows us to rotate\footnote{The MSSM Lagrangian and the soft terms are not invariant under such a rotation. However, these terms are negligible during inflation and reheating. One can also think of the $SU(3)$ rotations that we perform as a change of basis for the chiral fields $L^a$. Such a change of basis is of course unavoidable, if we want to diagonalize the mass matrix.}
\be
L^a \rightarrow {U^a}_b L^b, \quad \text{with ${U^a}_b\in SU(3)$, such that} \quad p_a \rightarrow p_a {U^a}_b = \lp \begin{array}{c} p \\ 0 \\ 0 \end{array}\rp \in \mathbb{R}\,.
\ee
Furthermore, we have enough freedom left in ${U^a}_b$ to also take
\be
q_a \rightarrow q_a {U^a}_b = \lp \begin{array}{c} q_1 \\ q_2 \\ 0 \end{array}\rp\,,
\ee
with $q_1 \in \mathbb{C}$ and $q_2 \in \mathbb{R}$.

The mass matrix for the matter fields $\{L^1, L^2, L^3, H^u, \bar L^{\bar 1}, \bar L^{\bar 2}, \bar L^{\bar 3}, \bar H^u\}$ at the critical point $L^a=H^u=0$ during inflation then takes the form
\be
{\cal M}^2 = \lp V+|A|^2 \rp  \mathds{1} +  \lp\begin{array}{cccccccc}
p^2 |Z|^2 &0&0&0 &0&0&0& c_1\\
0&0&0&0&0&0&0&c_2\\
0&0&0&0&0&0&0&0\\
0&0&0&p^2 |Z|^2&c_1&c_2&0&0\\
0&0&0&\bar c_1&p^2 |Z|^2&0&0&0\\
0&0&0&\bar c_2 &0&0&0&0\\
0&0&0&0&0&0&0&0\\
\bar c_1 & \bar c_2 &0&0&0&0&0&p^2 |Z|^2\\
\end{array}\rp\,,
\label{Timm}\ee
where 
 \begin{align}
  c_1 = B(Z) q_1 - p \lp Z A(Z) -\frac{(1-Z^2)^2 A'(Z)}{3 \alpha} \rp \,, \qquad c_2 = B(Z) q_2 \,.
 \end{align}
The first term in the mass matrix denotes the scalar potential $V$ and is the standard Hubble induced mass $3H^2$ for scalar fields. The non-diagonal part of the mass matrix can be trivially reduced to a $6\times6$ matrix since the complex field $L^3$ is not getting an additional mass term apart from the 1st term which gives $L^3$ a mass of $V+|A|^2$. In the next section we illustrate the physical consequences of this mass matrix by focussing on a specific model.

\section{Stability and  post-inflationary evolution}

In this section we will analyze the (post-)inflationary evolution and the appearance of instabilities in a specific model example. In particular, we will neglect the field $L^3$ that is decoupled during inflation and we will treat $L^a$ and $H^u$ as $SU(2)$ singlets. Secondly, we take the specific superpotential
\be
W=  M S Z  +  \Big (p \, Z  L^{1} + q \, S L^2\Big)   H^u   \, ,
\label{KhalfNew2}
\ee
with $p,q \in \mathbb{R}$. This results in a number of simplifications since $A(Z) = 0$ and $c_1=0$. This superpotential is identical to the sneutrino inflation superpotential in eqn. (2.4) of \cite{Nakayama:2016gvg} after using their equations (2.13) and (2.14). Since the non-trivial \K\ geometry of our model is important only during inflation, all conclusions concerning the post-inflationary evolution in our model to be discussed below will remain valid for the model discussed in \cite{Nakayama:2016gvg}.

Looking at the mass matrix in eqn. \eqref{Timm} we see that this means that we are removing the mixing between $L^1$ and $H^u$; the mass squared for $L^1$ is then simply given by $(M^2+p^2) Z^2$. The rest of the mass matrix reduces to two $2\times2$ blocks with eigenvalues
\be\label{eq:eigen}
\mu^2_\pm = M^2 Z^2 + {1\over 2} \Big (p^2 Z^2 \pm \sqrt{ 4 M^2 q^2 Z^2 + p^4 Z^4}\Big )\, .
\ee
The potential has thus two potentially tachyonic directions due to the mixing of the $L^2$ and $H^u$. As mentioned above, this is expected since we do not satisfy the stability criteria derived in \cite{Kallosh:2016ndd}. 
In this class of models, one typically has $p \sim q \sim O(1)$  \cite{Nakayama:2016gvg}, whereas $M^{2} \sim 10^{{-10}}$ for a proper normalization of the spectrum of inflationary perturbations.
By expanding the lower mass eigenvalue in powers of the small parameter $M$, one finds the following simple expression valid for $Z \gg M \sim 10^{-5}$:
\be
\mu^{2 }_- =  M^{2}\left(-{q^{2}\over p^{2}}+  Z^{2} \right) + \mathcal{O} (M^4) \ .
\label{eigen2}\ee
Since the hyperbloc geometry restricts $|Z| < 1$, this result implies that for $q \geq p$ there are two directions which are always tachyonic. This is true, in particular, for the typical case $q = p$ of \cite{Nakayama:2016gvg}.

To make this tachyonic direction explicit, let us look at the potential of the fields $L^{2}$ and $H^u$ along the inflationary direction ${\rm Re}\, Z$ (setting all other fields to zero since they are stable)
\be
V = e^{|L^{2}|^{2}+|H^u|^{2}}\, \Big( |MZ+ q L^2 H^u|^{2}  + |p Z H^u|^2\Big)\geq 0 \ .
\label{pot}\ee
Note that it is manifestly non-negative and has a stable Minkowski vacuum when both terms vanish separately. However, during inflation the first square does not vanish due to the non-vanishing of the inflaton field, leading to a potential instability. 

In terms of real components, 
 \begin{align}
 L^{2} = Z\, a\, e^{i\theta} \,, \qquad H^u =M\, h\, e^{i\omega} \,, 
 \end{align} 
the potential is given by
\be\label{modpot}
V  =  M^{2}Z^{2}\, e^{Z^{2}  a^{2}+M^{2}h^{2}}\Big(1  +   h^2 \bigl(p^{2}+q^{2} a^{2} \bigr)  + 2 q\,  h\, a\, \cos (\theta+\omega) \Big)\,.
\ee
Note that here we effectively rescaled the slepton and Higgs fields $L^{2}$ and $H^u$ in this expression: The field  ${L^{2}\over Z} =  a e^{i\theta}$ is given in the  units of the inflaton field $Z$, which starts at $Z \approx 1$ during inflation, and then gradually decreases. Meanwhile the rescaled Higgs field ${H^u\over M} =h e^{i\omega}$ is expressed in units of the inflaton mass $M \sim 10^{{-5}}$. 

The extremum of the potential with respect to $\theta$ occurs at $\theta+\omega = n\pi$; without any loss of generality one can take $\theta+\omega = \pi$. Then the potential becomes
\be\label{modpot2}
V  =  M^{2}Z^{2}\, e^{Z^{2}  a^{2}+M^{2}h^{2}}\Big(1  +   h^2 \bigl(p^{2}+q^{2} a^{2} \bigr)  - 2 q\,  h\, a\,  \Big)\,.
\ee
The last term reveals the only potential source of the instability: a possibility of a simultaneous growth of $h  = {|H^u|\over M}$ and $a={|L^{2}|\over Z}$.  This  corresponds to  the eigenvector with the possibly tachyonic mass eigenvalue $\mu_-^2$.

There are two ways to partially remedy this problem. First of all, one can add the stabilizing term $- \zeta  S\bar S  L^{2} \bar L^{2}$ to the \K\ potential  \cite{Nakayama:2016gvg}. This modifies the expression for the lowest mass eigenstates:
\be
\mu_-^{2 } =  M^{2}\left(-{q^{2}\over p^{2}}+  (1+\zeta) Z^{2} \right) + \mathcal{O} (M^4) \ .
\ee
This correction helps for sufficiently large $Z$: during inflation, with $Z\approx 1$, we require that\footnote{Note the close similarity of this condition to the constraint in eqn. (2.16) in \cite{Nakayama:2016gvg}:
\be
\mu_-^2 = kH^2 - \frac{q_a \bar q^{a}}{p_a \bar p^a} M^2 > 0 \ , \qquad k \in \mathbb{N} \,.
\ee}  
\be
1 +\zeta  > {q^2\over p^2}  \ .
\ee
However, after inflation  the field $Z$ decreases, and one can see from \rf{eigen2} that the tachyon instability develops for 
\be
Z^2< {q^2\over (1+\zeta) p^2} \ .
\ee 
Until this happens, the \K\ stabilization works pretty well. However, as soon as the tachyonic instability develops, a detailed investigation of the behavior of the system becomes unreliable  because the \K\ stabilization  does not protect the fields $L^{2}$ and $H^{u}$ against falling towards the boundary of the moduli space. 
This can be seen by following the \K\ metric of the stabilizer field along the tachyonic direction, which can be approximated by
 \begin{align}
  K_{S \bar S} = \frac{1}{1 - \frac{p^2}{q^2} \zeta Z^2} - \zeta  + \mathcal{O}(M) \,.
 \end{align}
Therefore any significant modification of the \K\ potential with $\zeta > 1$ leads to a tachyonic instability that renders the stabilizer fields ghost-like at some finite $Z$.

 One could expect that this instability must stop at sufficiently large values of the fields $L^{2}$ and $H^{u}$, where the potential develops an infinitely high barrier due to the stabilizing terms. However, a detailed investigation of this issue shows that the minimum of the stabilized potential occurs infinitesimally close to the boundary of the moduli space, where the potential barrier becomes infinitesimally thin. In this situation one cannot be sure that the \K\ stabilization used in  \cite{Nakayama:2016gvg} actually protects the stability of the system once the tachyonic instability started to develop.

Fortunately, one can avoid this problem.  A similar modification of $\mu_-^{2}$ can be achieved  by replacing the constant $q$ in the superpotential by $q(1-\beta Z^{2})$, and replacing $p$ by $p(1+\gamma Z^{2})$.  Since these extra terms disappear  at the minimum of the potential at $Z = 0$, this modification does not affect the late-time neutrino mass matrix. During inflation, however, the mass is modified to
\be \label{tm}
\mu_-^{2 } =  M^{2}\left( Z^2 - \frac{q^{2} (1 -\beta Z^{2})^{2}}{p^2 (1 + \gamma Z^2)^2} \right) + \mathcal{O} (M^4) \ .
\ee
This leads to a stabilization of the lepton and Higgs fields if $\beta$ and $\gamma$ take sufficiently large positive values. For instance, taking $p=q=1$ and $\gamma =0$, the instability is absent as long as we satisfy
\be\label{phtr} 
Z^2(1+2\beta - \beta^2 Z^{2}) > 1  \ .
\ee 
Just as with the \K\ stabilization, this method works only until the field $Z$ becomes small. One can delay the onset of the instability by increasing $\beta$, but it is extremely difficult to avoid it.  This instability leads to spontaneous symmetry breaking with generation of the classical fields $L^2$ and $H^u$, but without the problematic issues with the boundary of the moduli space, which emerge as a result of the \K\ moduli stabilization.

\begin{figure}[t!]
\centering
{
\includegraphics[width=\textwidth]{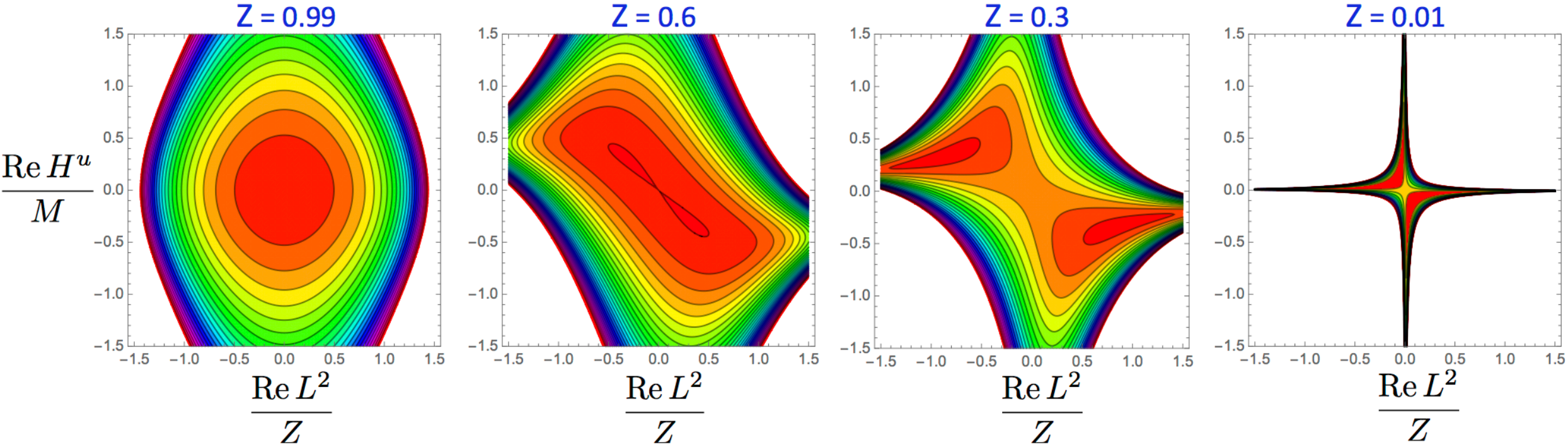}
\vskip - 20pt
}~~~~~
\caption{\label{ff1}\footnotesize  The contour plot of the potential \rf{pot}  divided by $M^{2}Z^{2}$, as a function of the rescaled real part of the slepton field  $ {{\rm Re}\,L^{2}\over Z}$ and a rescaled real part of the Higgs field ${{\rm Re}\, H^{u}\over M}$. The potential is stabilized by using $q = p (1 - Z^2)$. The dark red area corresponds to the minimum of the potential. The first figure corresponds to  $Z =0.99$, describing very early stages of inflation. The second figure is for $Z = 0.6$, where the tachyonic direction appears and which for $\alpha=1$ corresponds to approximately two e-foldings before the end of inflation. The last two figures correspond to post-inflationary values $Z = 0.3$ and $Z = 0.01$, where one sees the two flat directions emerging. Note that the Higgs rescaling means that the true value of the imaginary part of the Higgs field is five orders of magnitude smaller.}
\end{figure}

We illustrate this statement for the simplest case with $q = p$, $\gamma=0$ and $\beta = 1$ in  Fig. \ref{ff1}. With this simple choice of parameters, the negative term in \rf{tm} almost exactly disappears during inflation, which occurs at $Z \approx 1$. As a result, during the main part of inflation at $Z \approx 1$ the mass eigenvalue is positive. In accord with \rf{phtr}, there is no tachyonic instability  during inflation until the last two e-foldings of inflation when the field $Z$ becomes smaller than $(\sqrt 5-1)/2 \approx  0.618$.   In that case, the emerging tachyonic instability does not fully develop until well after the end of inflation.  One can stabilize the fields even further, by considering $\gamma > 0$ in \eqref{tm}.  \footnote{The inflaton dependent coupling to light moduli may lead to an additional contribution to inflationary perturbations \cite{Dvali:2003em}. We are grateful to Tomohiro Fujita who observed that, in general, one may encounter this in our scenario. Our stabilization mechanism \rf{tm}, \rf{phtr} is designed to avoid generation of these additional perturbations until  the very end of inflation. However, such additional perturbations might appear if we turn off the stabilization.}

\begin{figure}[t!]
\centering
{
\includegraphics[width=8cm]{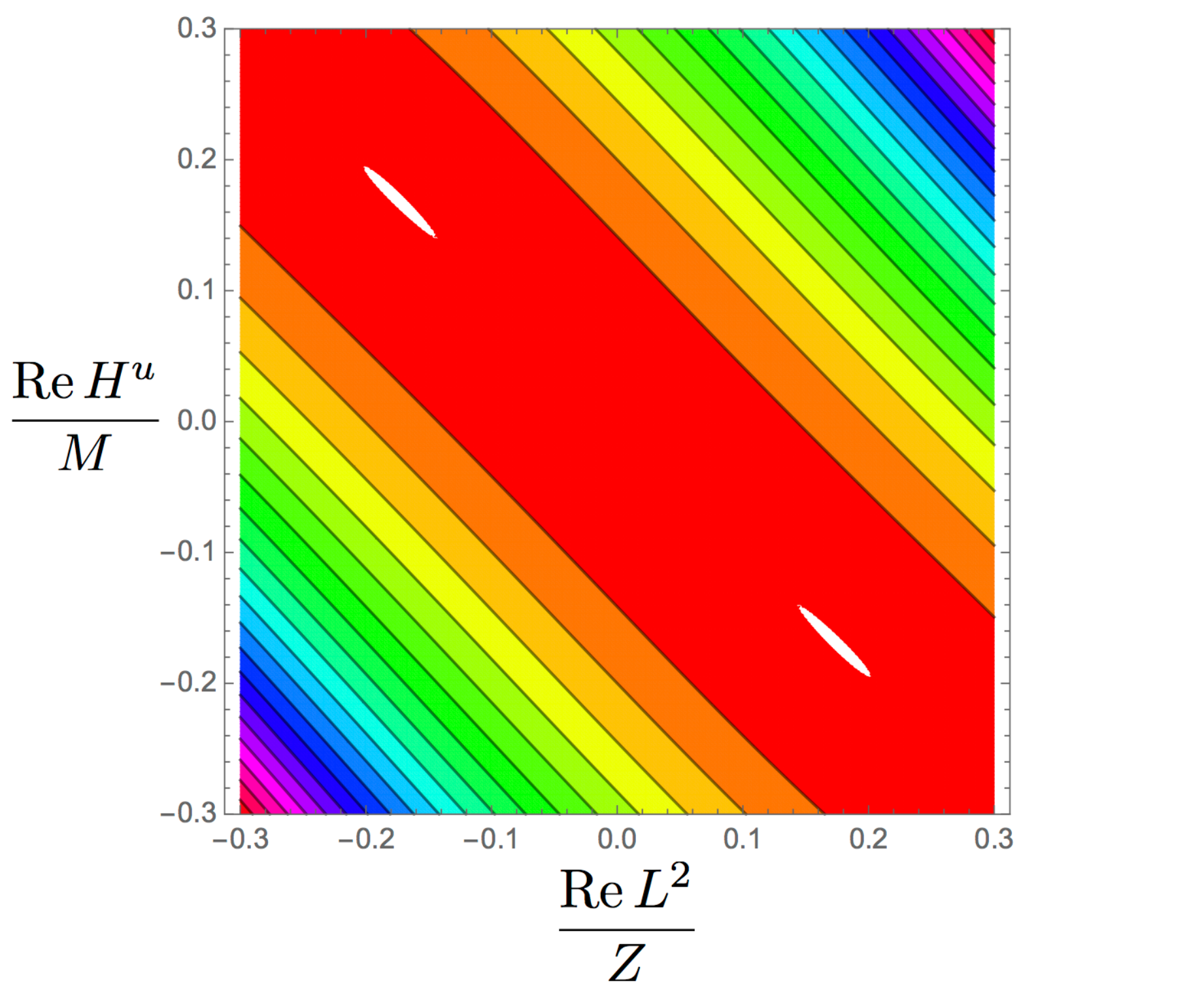}
}~~~~~
\caption{\label{ff2}\footnotesize The contour plot of the  potential \rf{pot} as a function of the rescaled real part of the slepton field  $ {{\rm Re}\,L^{2}\over Z}$ and a rescaled real part of the Higgs field ${{\rm Re}\, H^{u}\over M}$ for $Z = 0.985$ and $q = p$, which corresponds to the case without stabilization. The white spots correspond to the two minima of the potential. The value of the potential at  ${\rm Re}\,L^{2} =  {\rm Re}\,H^{u} =0$ is  $V = M^{2}Z^{2}$. The value of the potential at each of the two minima is   $0.9996\,  M^{2}Z^{2}$, i.e. it is different only by $4 \cdot 10^{-4} \,  M^{2}Z^{2}$.}
\end{figure}

In general, one may wonder whether we need any stabilization at all, since  we already learned that it can only temporarily protected the from the tachyonic instability. One can obtain an analytic understanding of the development of this instability from the scalar potential \eqref{modpot2}, without using any stabilization. We will be interested in $Z \gg M$ and hence neglect the second term in the exponential. In this approximation, the minima for the matter fields are located at
 \begin{align}
  qh = a Z^2 \,, \quad a = \sqrt{\frac{1}{Z^2} - \frac{p^2}{q^2}} \,. \label{tachyon}
 \end{align}
Note that there is only an instability at $Z=1$ when assuming that $q$ exceeds $p$, rendering the square real; in other cases the origin is stable for matter fields. At these minima, the scalar potential along $Z$ becomes for $p=q$
 \begin{align}
 V = M^2 Z^4 e^{1 -Z^2} = M^2 Z^2 \bigl(1 -2 (1-Z)^2 + \ldots \bigr) \,.
 \end{align}
The instabilities therefore have a negligible effect on inflation; the scalar potential only decreases by the square of the distance to the boundary $1-Z$. For instance, at $Z = 0.985$ which for $\alpha = 1$ corresponds to the point $50$ e-foldings until the end of inflation, the potential has only lost $4 \cdot 10^{-4}$ of its height as compared to its height $M^2 Z^2$ at $H^{u} = L^{2} = 0$, as illustrated in Fig. \ref{ff2}. Thus, the tachyonic instability at this stage, even if it brings the system to one of these two minima,  practically does not change the height of the inflaton potential, and thus does not affect much the dynamics of inflation. A similar figure for $Z = 0.86$, corresponding to the last $10$ e-folds of inflation, shows the potential with two minima which are only only 4\% below the potential at the inflationary ridge.

A detailed theory of the development of the tachyonic instability (tachyonic preheating) is described in  \cite{Felder:2000hj}. Here we will only give its simple interpretation: Initially,  $ a = h = 0$, and the first derivative of the potential in the directions of these fields vanish. Therefore the classical equations of motion for these fields allow them to stay in this unstable state forever. However, quantum fluctuations of these fields with momenta smaller than the tachyonic mass $\mu_-$ grow exponentially, approximately as $e^{|\mu_-| t}$. The initial amplitude of quantum perturbations with momenta smaller than $|\mu_-|$ is $\delta\phi \sim {|\mu_-| / 2\pi}$. (For inflationary experts: This estimate is similar to the estimate for an average magnitude of inflationary perturbations  $\delta\phi \sim {H\over 2\pi}$.) The instability continues until ${|\mu_-|\over 2\pi}\, e^{|\mu_-|t}$  becomes comparable to the value of the field in the minimum of the potential, which is $O(1)$ in our model. For the typical tachyonic mass scale $|\mu_-| \sim M  \sim 10^{{-5}}$, the  time required for spontaneous symmetry breaking in our model is  
\be
\Delta t_{\rm sb}  \sim 10 M^{{-1}} \ .
\ee
In reality, it is even somewhat longer. First of all, $\mu_-^{2}$ approaches its asymptotic value $-M^{2}$ only gradually, at $Z\ll 1$. In addition,  the expansion of the universe slows down the motion of the field. Therefore the growing quantum fluctuations of the fields $ a = h  = 0$ remain extremely small; their effect on other fields remains negligible during the interval $\Delta t_{\rm sb}  \sim 10 M^{{-1}}$.

This time should be compared with the typical time intervals in our model. For example, during inflation, important changes may occur within the time interval corresponding to a single e-folding of inflation, 
\be\label{d1}
\delta t_{e} \sim H^{-1} \sim \sqrt 3 \, M^{{-1}} \ , 
\ee
whereas after inflation the efficiency of reheating and preheating is usually determined by the number of oscillations of the inflaton field proportional to the inverse inflaton mass,  
\be
\label{d2}
\delta t_{\rm osc} \sim \sqrt {3\alpha}\, M^{{-1}} \ .
\ee

Comparison of these three time scales shows that for $\alpha = O(1)$ the tachyonic instability does not have enough time to develop within few e-foldings, or within a few oscillations of the inflaton field after inflation. Nevertheless, this instability may  significantly affect the process of reheating, by making the first, nonperturbative part of this process (preheating) much more efficient. Indeed, preheating  works by making quantum  fluctuations of the fields grow exponentially due to parametric resonance during the field oscillations \cite{Kofman:1994rk}. This effect becomes even more efficient due to tachyonic instabilities, which allow these perturbations to grow even faster \cite{Felder:2000hj}. 

Thus, in general we deal with a very complicated process, tachyonic preheating, which can be fully studied only using lattice simulations along the lines of  \cite{lattice} and \cite{Felder:2000hj}. The theory of this process differs dramatically from the naive theory of reheating ignoring the combination of the tachyonic instability and parametric resonance; in some cases the fields may roll towards the minima of their potentials and relax there very rapidly, within just one or two oscillations \cite{Felder:2000hj}. In our model, the main part of this process is pretty simple: The fields fall down towards the flat directions of the potential.

Indeed, at small $Z$ the potential  flattens out in some directions and at $Z = 0$ the potential for the other fields becomes very simple:
\be
V  =  q^{2}\ e^{|H^u|^{2} + |L^{2}|^{2}}\, |L^2|^{2} |H^u|^{2}  = q^{2}\ e^{e^{Z^{2}  a^{2}+M^{2}h^{2}}}\, M^2 Z^2 a^2 h^2 \,.
\ee
The potential is positive everywhere except along the two flat directions: It vanishes for all values of $L^{2}$, if $H^u = 0$, and for all values of $H^u $, if $L^{2} = 0$.   The emergence of these flat directions with a gradual decrease of the field $Z$ is apparent in  Fig. \ref{ff1}.

After the post-inflationary reheating of the universe, the potential will acquire large thermal corrections proportional to $T^{2}$, which will push the fields $H^u$ and $L^2$ along the flat directions towards zero \cite{KL72,Linde:2005ht}.  Moreover, the process of symmetry restoration begins well before the end of thermalization, due to the out-of-equilibrium particles produced during the preheating \cite{Kofman:1994rk,Kofman:1995fi}. Eventually, these fields should settle to their values determined by the potentials emerging in the  standard model and its generalizations.


The combination of all of these effects leads to the following scenario of the post-inflationary evolution: When the inflaton field $Z$ becomes sufficiently small, the system experiences spontaneous symmetry breaking with generation of classical fields $H^u$ and $L^a$ which may grow up to nearly Planckian values. However, in many cases the tachyonic instability may not have enough time to develop, and even if it does, it may be just a harmless (and useful) part of the scenario of tachyonic preheating. Like many processes in the theory of  preheating, the process of tachyonic preheating  may be over within a few oscillations of the inflaton field \cite{lattice,Felder:2000hj}. For example, the tachyonic preheating in the hybrid inflation scenario  \cite{Linde:1991km} may end very quickly, within a single oscillation of the inflaton field \cite{Felder:2000hj}.  

There is an interesting difference between the models that we studied in this paper and the hybrid inflation. In hybrid inflation, the scalar fields after the tachyonic waterfall acquire large vacuum expectation values which could lead to superheavy cosmic strings. Meanwhile we expect that in the models studied above  the values of the fields $H^u$ and $L^{a}$  after preheating and high temperature symmetry restoration should become very small, being determined by the small  mass terms  in the standard model defining the low-energy phenomenology of these fields. This is one of the issues which require a detailed phenomenological investigation.

\section{Discussion}
In this paper we  studied how to use the inflationary $\alpha$-attractor models to develop the idea of the minimal sneutrino inflation of the kind proposed in  \cite{Nakayama:2016gvg}, which was based on  supergravity implementation of the natural inflation scenario \cite{Kallosh:2014vja} with a discrete shift symmetry. In  \cite{Nakayama:2016gvg} the coupling of the inflaton to leptons and Higgs was arranged to depend on $\sin (n\vp/f)\leq 1$, so that it would never become very strong, even when the canonically normalized inflaton field $\vp$ takes super-Planckian values. We  argued that the same goal is easily reached in the context of the $\alpha$-attractor models since the coupling of  the inflaton to leptons and Higgs is arranged using the natural geometric Poincar\'e disk variable $Z=\tanh {\lp\vp / \sqrt{6\alpha}\rp}<1$. This stabilizes the behavior of masses and coupling constants during inflation,  making the inflaton coupling to matter exponentially small at large values of the inflaton field  \cite{Kallosh:2016gqp}. Yet another advantage of $\alpha$-attractors is that they provide one of the best fits to the existing observational data \cite{Ade:2015lrj}.
 
The attractive feature of the minimal sneutrino inflation proposed in \cite{Nakayama:2016gvg}, that is also a feature of our $\alpha$-attractor based models, is the possibility to realize the seesaw mechanism. Two heavy right-handed neutrinos are residing in the inflationary sector of the theory (inflaton and stabilizer superfields), and have characteristic couplings associated with the scale of inflation $M\sim 10^{-5}M_{Pl} \sim 10^{ 13}$ GeV.  
The light neutrinos then get the masses
\be
m_{ab} ^\nu = {v^2 \sin^2 \beta  \over M} \lp p_a p_b +q_a q_b\rp \,.
\ee
For $p_a,q_a\sim 1$ and $v^2\sim 10^{4} $ GeV we find the scale of the light neutrinos to be of the order $10^{4} \cdot 10^{-13}$ GeV which is of the order eV, where they should be.

In order to make this model minimal and to be able to describe the physics of neutrinos properly with regards to the MNS matrix, one has to make the inflaton sector interact directly with leptons and the Higgs. This takes the inflationary sector out of the hidden sector, which may lead to stability issues. It was recently shown in \cite{Kallosh:2016ndd} that when inflation is in the hidden sector, adding matter causes no tachyonic instabilities under certain conditions which are not very restrictive. It is particularly important for this purpose to avoid a direct interaction between matter fields and the stabilizer $S$, since this interaction may lead to tachyons.  Indeed, in agreement with \cite{Kallosh:2016ndd}, the negative terms in the mass eigenvalues in eq. \eqref{eigen2} are due to the coupling of the leptons and Higgs field to the stabilizer $S$. 

There are several different mechanisms which can be used to stabilize all fields  during inflation, so that the presence of matter  does not affect the inflationary evolution. 
The method proposed in \cite{Nakayama:2016gvg}  is based on adding stabilizing higher order terms  $- \zeta  S\bar S  L^{2} \bar L^{2}$ to the \K\ potential. We found that such terms can eliminate the tachyonic instability during inflation, but they become inefficient at small values of the inflaton field during the post-inflationary evolution.  Once the tachyonic instability develops, it may bring the fields dangerously close to the boundary of the moduli space, which makes investigation unreliable. We developed an alternative, simpler method of stabilization. Our mechanism does not eliminate the tachyonic instability, but it keeps it under control and renders it transient and harmless. This tachyonic instability becomes a part of the post-inflationary tachyonic preheating \cite{Felder:2000hj}.

A detailed description of this nonperturbative process in sneutrino inflation  requires lattice simulations similar to those performed in \cite{Felder:2000hj}, but we do not expect any surprises here: tachyonic preheating tends to stimulate the energy transfer from the inflaton to the matter fields, thus making reheating faster and even more efficient than the process considered in  \cite{Nakayama:2016gvg}. However, the process studied in  \cite{Nakayama:2016gvg} is already very efficient  because the interactions of the inflaton field with other matter fields is rather strong, which leads to a very high reheating temperature $T \sim 10^{14} - 10^{15}$ GeV.  The tachyonic preheating makes the whole process much more involved than the process studied in  \cite{Nakayama:2016gvg} which ignores the tachyonic instability, but this should not substantially alter the resulting estimate of the reheating temperature.  

In supergravity, a high reheating temperature is not necessarily a good thing, because a reheating temperature above $10^{8} $ GeV  may lead to the cosmological gravitino problem  \cite{Ellis:1982yb,GRAV,KKM}. One may solve this problem by considering a superheavy gravitino with mass $m_{3/2} \gtrsim 10^{2}$ TeV \cite{Nakayama:2016gvg,Dudas:2012wi,Ellis:2015jpg}. However, in that case one must  get rid of the excess of LSP produced by gravitino decay, e.g. by introducing R-parity violation  \cite{Nakayama:2016gvg}. Thus there is a price to pay for the high reheating temperature, which appears in sneutrino inflation, in Higgs inflation, as well as in any other inflationary model in supergravity in which the inflaton does not belong to the hidden sector.

On a positive side, a high reheating temperature requires a greater number $N$ of e-foldings of inflation, which slightly increases the spectral index $n_{s} = 1-2/N$ for $\alpha$-attractors. This may be an advantage of this class of models since it may further improve the compatibility of these models, as well as of the Starobinsky model, with the latest observational data \cite{Ade:2015lrj,Ellis:2015pla,Kallosh:2016gqp}.

Independently of all of these issues, we believe that the investigation of the tachyonic instability in this model provides an interesting example of a very unusual dynamical behavior where an instability may lead to the generation of scalar fields of nearly Planckian amplitude, which, however, tend to disappear after the subsequent cosmological evolution. Thus, instead of trying to avoid the tachyonic instability, one may try to learn whether one can use the new possibilities constructively. A similar waterfall tachyonic instability  was first discovered in the context of the hybrid inflation  \cite{Linde:1991km}. This effect resulted in the creation of superheavy cosmic strings, which ruled out some of the most popular versions of hybrid inflation in supergravity. In the new set of models discussed above, this waterfall regime can end up in a stable  supersymmetric Minkowski vacuum without producing any undesirable cosmological defects. One can break supersymmetry in this vacuum and uplift it to dS with a small cosmological constant without affecting investigation of inflationary evolution \cite{Linde:2016bcz}.

As we already argued in the Introduction, these results imply that a combination of cosmological attractors and vacuum stabilization in supergravity may significantly simplify the construction of realistic inflationary models involving a large number of scalar fields.

\section*{Acknowledgments}

We are grateful to S. Ferrara, T. Fujita, H. Murayama,  Y. Yamada and J. Yokoyama for enlightening  discussions.   The work of RK  and AL is supported by the SITP, and by the NSF Grant PHY-1316699.  The work of AL is also supported by the Templeton foundation grant `Inflation, the Multiverse, and Holography'. The work of TW is supported by COST MP1210. DR and TW thank the Department of Physics of Stanford University for the hospitality during a visit in which this work was initiated.



\end{document}